\documentclass[%
 reprint,
 amsmath,amssymb,
 aps,
floatfix
]{revtex4-2}

\usepackage{graphicx}
\usepackage{dcolumn}
\usepackage{bm}
\usepackage{xcolor}

\begin{document}


\title{Terahertz Metamaterial Renormalization of Superconducting Josephson Plasmons in La$_{1.85}$Sr$_{0.15}$CuO$_4$}


\author{Kelson Kaj$^1$}
\author{Ian Hammock$^1$}
\author{Chunxu Chen$^2$}
\author{Xiaoguang Zhao$^2$}
\author{Kevin A. Cremin$^1$}
\author{Jacob Schalch$^1$}
\author{Yuwei Huang$^2$}
\author{Michael Fogler$^1$}
\author{D. N. Basov$^3$}
\author{Xin Zhang$^2$*}
\author{Richard D. Averitt$^1$}
\email{Corresponding author email: xinz@bu.edu, raveritt@ucsd.edu}
\affiliation{
$^1$Department of Physics, University of California San Diego, San Diego, CA \\
$^2$Department of Mechanical Engineering, Boston University, Boston, MA\\
$^3$Department of Physics, Columbia University, New York, NY
}

\begin{abstract}
 We investigate light-matter coupling in the cuprate  superconductor La$_{1.85}$Sr$_{0.15}$Cu0$_4$ (LSCO), accomplished by adhering metamaterial resonator arrays (MRAs) to a c-axis oriented single crystal. The resonators couple to the Josephson Plasma Mode (JPM) which manifests as a plasma edge in the terahertz reflectivity in the superconducting state. Terahertz reflectivity measurements  at 10K reveal a renormalization of the JPM frequency, $\omega_{jpm}$, from 1.7 THz for the bare crystal to $\sim$1 THz with the MRAs. With increasing temperature, the modified $\omega_{jpm}$ redshifts as expected for decreasing superfluid density, vanishing above T$_{c}$.  The modification of the electrodynamic response arises from resonator induced screening of the longitudinal polariton response, reminiscent of plasmon-phonon coupling in doped semiconductors. Modeling reveals that the electrodynamic response is fully interpretable using classical electromagnetism. Future studies will have to contend with the large effects we observe which could obscure subtle changes that may indicate cavity-based  manipulation of superconductivity. Finally, we note that our MRA/LSCO structure is a tunable epsilon-near-zero (ENZ) metamaterial that exhibits a nonlinear response arising from the c-axis Josephson tunneling coupled with the local fields of the resonators.

\end{abstract}

\pacs{Valid PACS appear here}
\maketitle

\section{Introduction}
The use of light to study and manipulate materials has led to many interesting and important developments in the study of quantum materials, such as photoinduced phase transitions, tunable devices, and light-enhanced phases \cite{Basov2017,Schiffrin2013,Hu2014,Zhang2014,Cavalleri2018,Yang2023,Bloch2022,delaTorre2021}. The incorporation of cavities with quantum materials has further enhanced the ability of researchers to couple light with novel phases and collective modes of quantum materials \cite{Schlawin2022,FornDiaz2019,Bloch2022,Ahn2021,Hubener2021,Nagarajan2021,Li2018,LiDicke2018,Jarc2023,Thomas2021,Baydin2023}. The integration of cavities with quantum materials provides a potential route to modify ferroelectric phase transitions, enhance superconducting correlations, or induce topological switching \cite{Ashida2020,Latini2021,Laplace2016,Hubener2021,Appugliese2022}. Integrating cavities with superconductors is a particularly intriguing possibility as it has been shown that cavities can be used to cool superconducting phase fluctuations, and calculations suggest that coupling strengths and field strengths necessary to accomplish this are achievable  \cite{Laplace2016,Fistul2007,Hammer2011}. However, experimentally realizing cavities integrated with quantum materials is a challenging task, further complicated by the fact that cavities fabricated directly with quantum materials may require the growth of multiple samples to perform a frequency sweep of the cavity resonance to clarify or optimize the coupling.

In this work, we study the interaction between cavities and excitations in quantum materials by integrating metallic metamaterial resonator arrays (MRA) with a superconducting cuprate through coupling to the Josephson Plasma Mode (JPM) which occurs in the THz frequency range. Our primary goal is to investigate, experimentally and theoretically, the electrodynamic response of metamaterial resonator arrays (MRAs) integrated with the c-axis cuprate La$_{1.85}$Sr$_{0.15}$Cu0$_4$ (LSCO) with a view towards cavity augmented superconductivity. The present results build upon our previous work showing the coupling between a metamaterial resonance and the JPM in LSCO \cite{Schalch2019}. Improved fabrication of the MRAs enables changing the resonator array-LSCO surface spacing. We investigate the terahertz response for arrays that are placed $\sim$200 nm and $\sim$9 $\mu m$ from the LSCO surface. Both samples reveal a renormalization of the JPM frequency, $\omega_{jpm}$, from 1.7 THz to 1.1 THz at 10K. The observed shift is consistent with the appearance of the lower superconducting polariton arising from strong electrodynamic coupling between the resonators and the JPM with somewhat stronger coupling for 200 nm spacing in comparison to 9 $\mu m$. More specifically, the modification of the electrodynamic response arises from resonator induced screening of the longitudinal LSCO response, reminiscent of plasmon-phonon coupling in doped semiconductors \cite{Yokota1961,Mooradian1966,Mooradian1967,Tell1968}. With increasing temperature, the hybrid LSCO/MRA mode redshifts as expected for a decreasing superfluid density. The data can be explained using classical electromagnetic coupling with no evidence of cavity-based modification of the microscopic parameters that determine the superconducting response of LSCO. Nonetheless, our results demonstrate strong light-matter coupling as a step towards optimizing the ability to tune and manipulate superconductivity using terahertz cavity-based approaches.

\begin{figure}[t]
    \centering
    \includegraphics[width=5cm]{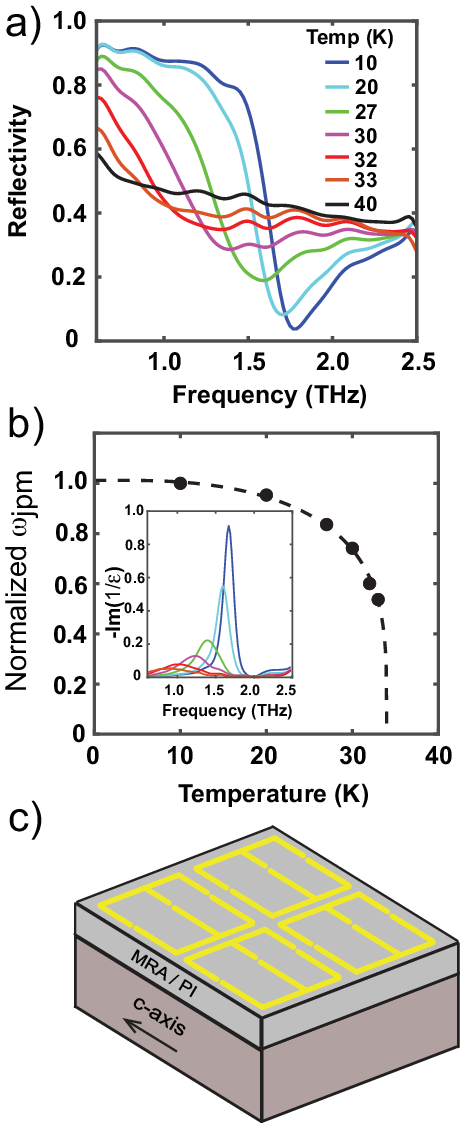}
    \caption[THz reflectivity of LSCO with loss function and LSCO-MM schematic]{a) Reflectivity of LSCO crystal. b) JPM frequency vs. temperature normalized to the 10 K value, inset: loss function extracted from LSCO reflectivity. c) Schematic of MRA tape placed on a c-axis oriented LSCO crystal.}
    \label{fig:lscommfig1}
\end{figure}

Superconducting cuprates are comprised of layers of quasi-2-d superconducting copper-oxygen planes. In the normal state ($T > T_c$, where $T_c$ is the superconducting transition temperature), cuprates are insulating along the c-axis. However, in the superconducting state, transport along the c-axis arises from tunneling of Cooper pairs between Cu-O planes due to the Josephson effect \cite{Josephson1962}. The tunneling current is directly related to the order parameter phase difference between the layers, which can be modified by an electric field \cite{Savelev2010}. The $c$-axis plasma frequency associated with this transport is the JPM, and it manifests in the reflectivity as a plasma edge \cite{Tamasaku1992,Kaj2023,Laplace2016a,Dordevic2003}. For example, the reflectivity of the La$_{1.85}$Sr$_{0.15}$Cu0$_4$ (LSCO) single crystal studied here is shown in Fig. \ref{fig:lscommfig1}a.  

Given that the JPM is a plasma frequency, it is a longitudinal excitation and shows up as a peak in the loss function (defined as $-1/Im(\epsilon))$, as shown in the inset of Fig. \ref{fig:lscommfig1}b. The JPM frequency is proportional to the square root of the superfluid density. Thus, as the sample is cooled further into the superconducting state, the plasma edge blueshifts to higher frequencies, following the superconducting order parameter, as shown in Fig. \ref{fig:lscommfig1}b. The LSCO single crystal studied here has a JPM of 1.7 THz at 10 K.

To study the interaction of this longitudinal electromagnetic feature with a cavity, we use metallic metamaterial (MM) resonators placed on top of the LSCO as depicted schematically in Fig. \ref{fig:lscommfig1}c. Metamaterials are periodic arrays of sub-wavelength structures where the resonance is controlled by the geometry of the resonators\cite{fan2022,Pendry2006,Schurig2006,Padilla2006,Padilla2007,Smith2000}. Instead of fabricating the metamaterial resonator array (MRA) directly onto the surface of the single crystal, the arrays are fabricated as free-standing polyimide tapes, and then placed on the surface of a c-axis oriented LSCO single crystal (Fig. \ref{fig:lscommfig1}c). The total thickness of the polyimide tape is 9.4 $\mu m$, and the position of the resonator layer within the tape can be controlled, thereby changing the distance between the resonators and the LSCO crystal surface.

\begin{figure}
    \centering
    \includegraphics[width=6.5cm]{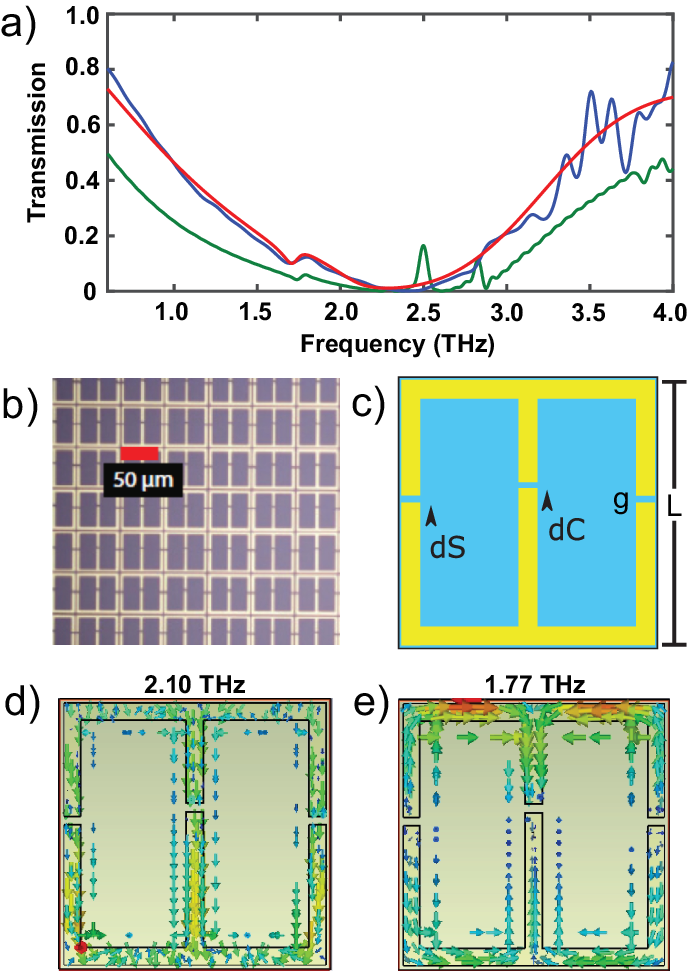}
    \caption[Metamaterial Tape Transmission, schematic, and simulated surface currents]{a) Isolated MRA tape transmission: Simulated (green), Experimentally measured (blue), and modelled with two Lorentzian oscillators (red). b) Microscope image of a portion of the MRA (scale bar 50 $\mu m$). c) Single MM unit cell - dimensions indicated by black arrows. The metamaterial dimensions are as follows:  gold bar length L=59 $\mu m$, gap value g=2 $\mu m$, center gap offset dC=6 $\mu m$, side gap offset dS=3 $\mu m$. The array periodicity is 61 $\mu m$ and the width of gold lines is 4 $\mu m$. Full-wave electromagnetic simulations of surface current: d) at dipole resonance frequency of 2.10 THz and e) at asymmetric resonance frequency of 1.77 THz.}
    \label{fig:lscommfig2}
\end{figure}

The geometric details and transmission of the free-standing tapes (i.e., prior to adhering to the LSCO crystal) are shown in Fig. 2. The resonators consist of two asymmetric ``E-shaped" gold split ring resonators with the geometric parameters labelled in Fig. \ref{fig:lscommfig2}c. A microscope image of  a portion of the array is shown in \ref{fig:lscommfig2}b. Full-wave electromagnetic simulations of the bare tape (i.e., not adhered to to the LSCO crystal) transmission (green curve) and the experimentally measured transmission (blue curve) are shown in Fig. \ref{fig:lscommfig2}a. In addition, the calculated transmission of the tape modelled with two Lorentzian oscillators is plotted (red curve). The surface currents for the broad dipole mode at 2.10 THz and the high-Q asymmetric 1.77 THz are shown in Fig. \ref{fig:lscommfig2}d and \ref{fig:lscommfig2}e, respectively. The 2.10 THz dipolar mode is characterized by parallel currents in each of the dipole bars. The asymmetric mode at 1.77 THz exhibits out-of-phase currents both between the parallel bars and between the 2 asymmetric ``E" shaped split ring resonators. As shown previously, the asymmetric mode is associated with a large field enhancement in the center gap \cite{Chen2022}. In the present case, the asymmetric mode provides an additional marker to track the electromagnetic response when the MRA tapes are adhered to the LSCO. When placed on the LSCO crystal (as shown in Fig. 1c and describe in detail below), the dipole bars are parallel to the c-axis, as is the incident THz electric field.

For the effective medium modeling below, we treat the MRA tape as having two Lorentzians corresponding to the dipole and asymmetric modes, given by the following equation:

\begin{equation}
    \label{eq:mmepsilon}
    \epsilon_{MM}(\omega)=\epsilon_{PI}+\sum_i^{1,2} \frac{F_i}{\omega_{MMi}^2-\omega^2-i\omega\gamma_{MMi}}
\end{equation}

\noindent In this equation, $\epsilon_{PI}$ is the polyimide dielectric constant, $F_i$, $\omega_{MMi}$ and $\gamma_{MMi}$ are the oscillator strength, resonant frequency and damping of the $i^{th}$ MM resonances ($i=$ 1, 2). The parameters used for the dual Lorentzian models are given in table S1 of the Supplementary Information (SI). The dipole resonance has a large oscillator strength and large damping, giving the broad downward slope in the transmission in Fig. \ref{fig:lscommfig2}a \cite{Chen2022,Cong2019}. The asymmetric resonance is much sharper with a weaker oscillator strength, giving the peak-dip feature in the THz transmission at $\sim$1.75 THz in Fig. \ref{fig:lscommfig2}a. Modeling using Eqn. 1 yields the red curve for the transmission of the MRA tape in Fig. 2a. It provides a good fit to the data (blue curve). In short, the data of the bare LSCO response (Fig. 1) and the bare MRA tape (Fig. 2) provide a starting point to understand the electrodynamic response of the integrated LSCO/MRA structures presented below.

\section{Materials and Methods}
The metamaterial tapes were fabricated with conventional surface-micromachining. First, the polyimide film (PI-2600 series, HD MicroSystems) was coated onto a silicon wafer at a specified thickness. The polyimide thin film was then cured using rapid thermal annealing (RTA) in nitrogen. The ramp up speed is 3 $^{\circ}$C/min and is held at 350 $^{\circ}$C for 30 minutes. For the thinner polyimide layers (i.e., 200 nm) the thicker film was etched with reactive ion etching (RIE) with an etching rate of ~150 nm/min with 10 sccm CF4 and 40 sccm O$_{2}$ under 200 mTorr and 300 W plasma environment. Then the gold metamaterial structures were defined through photolithography and lift-off processes. A second polyimide film was coated on top of the metamaterial structures followed by the same curing process described above. Finally, the polyimide film was peeled off from the silicon substrate. The metamaterial tapes were mounted onto the LSCO surface with the dipole bars pointing along the c-axis of the LSCO crystal.

THz measurements were performed in an evacuated, home-built vacuum box.  The THz was generated using a portion of the output of a 1 KHz, 6 W laser system with a ZnTe crystal. After reflecting from the sample surface, the THz was measured using electro-optic sampling (EOS) in another ZnTe crystal with an 800 nm gate beam. A gold mirror was used as the reference. Additionally, high-field THz measurements were performed using a standard tilted pulse front LiNbO$_3$ based setup, achieving THz fields up to 80 kV/cm \cite{Hebling2004,Hebling2002,Hirori2011,Shimano2012}. The LSCO crystal was mounted onto a continuous flow cryostat and cooled with liquid helium. The ZnTe generation and detection crystals are in the vacuum chamber thereby eliminating drift issues associate with water vapor. The LSCO crystal was cut to expose the a-c plane and has a doping of approximately 15$\%$, with a JPM of $\sim$1.7 THz, agreeing with previous measurements \cite{Tamasaku1992,Kaj2023,Dordevic2003}.

\section{Results}

The experimental results and comparison with full-wave electromagnetic simulations of the coupling of the MRA and the c-axis superconducting response are presented in Fig. \ref{fig:lscommexpt}. However, before presenting the experimental results, we first investigate the electromagnetic response using a simple analytical effective medium analysis and an analytical multilayered model. These results are presented in Fig. \ref{fig:lscommcalcs}. The effective medium model highlights the longitudinal polaritonic nature of the electromagnetic coupling that arises from screening, leading to well-defined upper and lower polariton branches. However, more realistic multilayer modeling reveals that the upper polariton does not clearly manifest in the reflectivity. These analytical considerations set the stage for understanding the experimental results and full wave simulations in Fig. \ref{fig:lscommexpt}.

\subsection{Effective Medium and Fresnel Multilayer Analysis}

\begin{figure*}
    \centering
    \includegraphics[width=17cm]{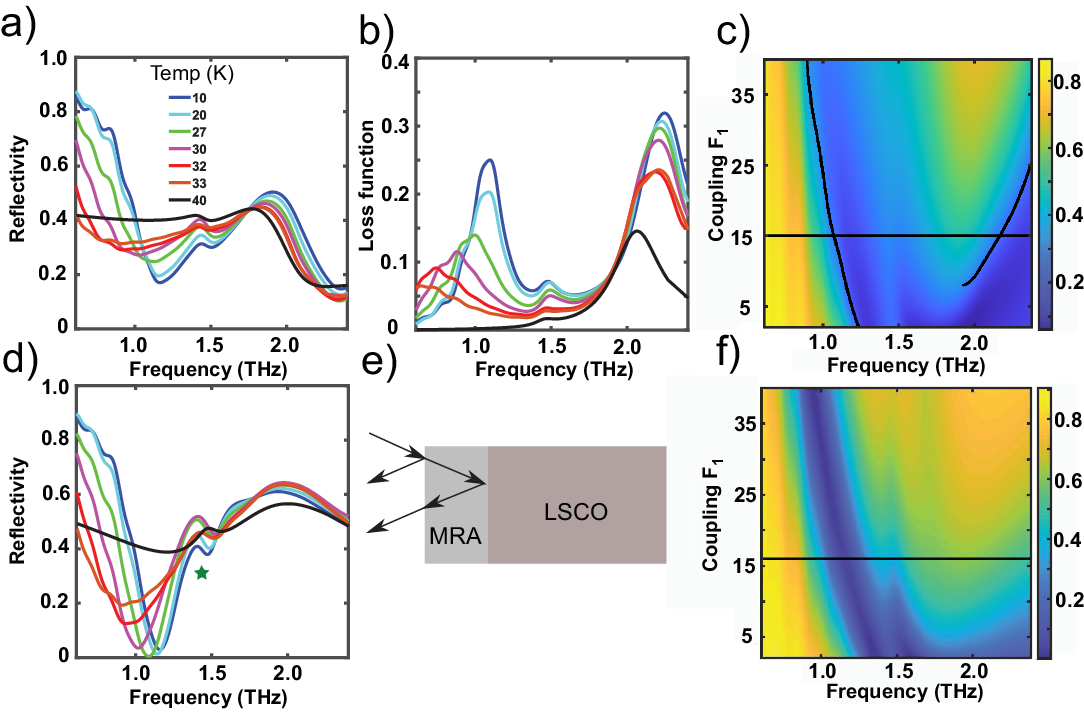}
    \caption[Calculated Reflectivity for LSCO-MM with Single-Medium and Interference Theory]{a-b) Effective medium theory calculation of reflectivity (a) and loss function (b) as a function of temperature, using the dielectric function from Eqn. \ref{eq:lscommeffmed} with LSCO parameters extracted from the experimental reflectivity, and MRA parameters (see Table 1 in SI). c) 2-d color plot of reflectivity calculated using the dielectric function in equation \ref{eq:lscommeffmed} as a function of coupling strength $F_{1}$ at 10K. d) Calculation of reflectivity as a function of temperature using the multilayered interference theory model. e) Schematic of interference theory model, where MM-tape is modelled as a film on top of bulk LSCO. f) 2-d color plot of reflectivity calculated using multilayered interference theory modeled as a function of coupling strength $F_{1}$ at 10K.}
    \label{fig:lscommcalcs}
\end{figure*}

 In the simplest approach, the integrated MRA-LSCO sample is treated as an effective medium comprised of a Drude response for the LSCO JPM and Lorentzian oscillators for the metamaterial resonances \cite{Tamasaku1992}. This single-medium model has the following dielectric function:
 
\begin{multline}
\label{eq:lscommeffmed}
    \epsilon(\omega)=f \left ( \epsilon_{LSCO}-\frac{\omega_j^2}{\omega^2+i\omega\gamma}\right )+(1-f)\epsilon_{MM}(\omega) 
\end{multline}
\\
\noindent where $f$ is the volume fraction of the LSCO, $\epsilon_{LSCO}$ is the dielectric background of LSCO, $\omega_j$ is the JPM frequency, $\gamma$ the LSCO quasiparticle scattering rate, and $\epsilon_{MM}$ is the effective dielectric response of the MRA described in Eqn. 1.  For the calculations presented in Fig. \ref{fig:lscommcalcs}, the dielectric function extracted from the experimental reflectivity of LSCO in Fig. \ref{fig:lscommfig1}a is used in the modelling. The oscillator strengths of the metamaterial resonances $F_{1,2}$ (see Eqn. 1) describe the coupling as these  directly relate to the magnitude of the shift of the effective JPM response (i.e. lower polariton), as will be shown below. The MRA parameters used in the calculations are shown in Table S1 of the SI. The reflectivity calculated from Eqn. \ref{eq:lscommeffmed} for the 9 $\mu m$ MRA tape adhered to the LSCO is shown in Fig. \ref{fig:lscommcalcs}a.

The black curve in Fig. \ref{fig:lscommcalcs}a shows the reflectivity above T$_{c}$ (40 K). The features at $\sim$1.5 and $\sim$1.8 THz correspond to the MRA modes of the bare MRA (Fig. 2a), shifted from the enhanced local permittivity (i.e., $\epsilon_{LSCO}$ = 27). There is also an upper edge in reflectivity at $\sim$2.2 THz from the dipole resonance of the MRA. Below T$_{c}$, the calculated reflectivity exhibits reflectivity edges above ($\sim$2.3 THz) and below ($\sim$1.2 THz)  the bare LSCO JPM frequency (1.7 THz at 10 K). Both features redshift with increasing temperature, tracking the JPM which is proportional to the square root of the superfluid density. In contrast, the sharp asymmetric mode at 1.5 THz is relatively insensitive to the onset of superconductivity with decreasing temperature.

 Additional insight is gained from the calculated effective loss function, -Im(1/$\epsilon(\omega)$) shown in Fig. \ref{fig:lscommcalcs}b. The loss function identifies the zeros of $\epsilon(\omega)$ coresponding to the longitudinal response.  For the above - T$_{c}$ 40 K curve (black line), the loss function has  peaks at 2.0 and 1.5 THz corresponding to the longitudinal mode frequencies of the MRA when placed onto LSCO, with the higher frequency being the dipole resonance and the lower frequency that of the asymmetric resonance.  

The response at 40 K can be directly compared with the calculation at 10K (blue curve), where the loss function has peaks at 1.1, 1.5, and 2.3 THz. The two upper peaks that were present at 40 K are also present at 10K. The additional peak at 1.1 THz arises from the coupling between the JPM and MM modes and corresponds to the lower polariton branch with a shift of $\sim$30$\%$ from the bare JPM frequency. In addition, the upper peak of the loss function is shifted to higher frequency and corresponds to the upper polariton. Both modes exhibit the temperature response of the superconductor, redshifting as the temperature is increased due to a decrease in the condensate density. This shows that metamaterial-JPM coupling splits the JPM into upper and lower superconducting-polaritons (SCP). As with the reflectivity, the asymmetric peak in the loss function at 1.5 THz exhibits only small changes amplitude and frequency as the temperature is changed.

This coupling and shift of the longitudinal frequencies from MRA-JPM coupling is analogous to the shifts observed from longitudinal optical phonon-plasmon coupling in doped semiconductors \cite{Yokota1961,Mooradian1966,Mooradian1967,Tell1968} when the the doping is such that the plasma frequency is proximal to the optical phonon. For MRA-JPM coupling, the tuning results from temperature tuning of the condensate density. This provides an intuitive understanding of the mode coupling in terms of dynamic screening arising from the longitudinal character of the polar metamaterial modes, with the asymmetric mode being weakly coupled in comparison to the dipole mode (i.e., no splitting and weak temperature dependence of the peak).

To further clarify the electrodynamic response, the reflectivity calculated as a function of F$_{1}$ is shown in Fig. \ref{fig:lscommcalcs}c. The reflectivity edges can be seen as the transition from high reflectivity (gold) to low reflectivity (blue). and we can see these edges split from the bare JPM as F$_1$ is increased. The black lines in Fig. \ref{fig:lscommcalcs}c track the upper and lower peaks of the calculated loss function. Below a coupling F$_1$ of $\sim$10 the coupling is too small for the upper peak in the loss function to be resolvable. This indicates that oscillator-JPM coupling yields longitudinal upper and lower polaritons where the magnitude of the splitting is largely determined by F$_{1}$. The horizontal black curve is the value of $F_{1}$ that matches the lower polariton to the frequency of the reflectivity minimum seen in experiment, shown below in Fig. \ref{fig:lscommexpt}f.

In the absence of this analysis, it could be tempting to relate the coupling induced redshift of the lower polariton (at a given temperature, compared to the bare $\omega_{jpm}$) to a modification of the condensate density (as occurs for the bare JPM with temperature). However, our analysis shows that the renormalization of the polariton compared to the bare JPM is from electromagnetic coupling of the JPM to the metamaterials oscillators and the resultant change of the zero-crossing of the effective dielectric function. It does not arise from weakening of the superconductivity or oscillator induced change of the superfluid density. This is an important result that  highlights that any actual modification of the condensate will have to be disentangled from the electrodynamics that arise from polaritonic coupling. For example, a modification of the condensate density would manifest as a clear change of the superconducting transition temperature. We note that Fig. \ref{fig:lscommcalcs} a) – c) was presented to provide some insight into the coupling. However, it is a severe approximation, so we next consider an analytical multilayer model to more accurately capture the MRA-JPM coupling.

We now consider analytical calculations treating the MRA as a layer on top of the c-axis LSCO as in experiment (see Fig. \ref{fig:lscommcalcs}e). The MRA is taken as a 9.4 $\mu m$ thick planar array of Lorentzian oscillators on top of the bulk LSCO, with the dielectric function as in Eqn. 1. The reflectivity is analytically calculated using the following Fresnel equations, which includes multiple reflections inside of the MRA (but not in the thick LSCO crystal).
\begin{equation}
    r=\frac{r_{air,MM}+r_{MM,LSCO}e^{i2\delta}}{1+r_{air,MM}*r_{MM,LSCO}e^{i2\delta}}
    \label{eq:multilayerrefl}
\end{equation}
\begin{equation}
    \delta=2\pi d \sqrt{\epsilon_{MM}}\omega/c
\end{equation}
\noindent where $r_{air,MM}$ and $r_{MM,LSCO}$ are the Fresnel reflection coefficients for the air-MM and MM-LSCO interfaces respectively, $d$ is the thickness of the MRA, and $c$ is the speed of light. The MRA oscillator parameters with the MRA on the LSCO crystal are varied to match the frequencies of the features in the analytically calculated reflectivity with the experiment (Fig. \ref{fig:lscommexpt}). Specifically, this includes the lower polariton frequency and the peak-dip feature related to the asymmetric resonance. The results of the calculated reflectivity as a function of temperature are shown in Fig. \ref{fig:lscommcalcs}d. There is a clear reflectivity edge that redshifts with increasing temperature consistent with the effective medium model (previously ascribed to the lower polariton as discussed above). At 10 K this occurs at $\sim$1.15 THz, redshifed by 550 GHz from the bare JPM frequency. Additionally there is a peak-dip feature whose frequency is nearly temperature independent at $\sim$1.5 THz, as indicated by a green star in Fig. \ref{fig:lscommcalcs}d. This feature is at the transverse frequency of the asymmetric resonance, $\omega_{MM2}$. However, the reflectivity edge corresponding to the upper polariton is much broader, and does not show the clear temperature dependence as the effective medium model in Fig. \ref{fig:lscommcalcs}a. We note that for the multilayered calculation, a calculated loss function is not shown as the model is not characterized by a single dielectric function, thus making the extraction of a single loss function invalid.

The behavior of the calculated multilayer reflectivity as a function of coupling strength F$_1$ is shown in Fig. \ref{fig:lscommcalcs}f. We can see the effective lower polariton edge shifts to lower frequencies compared to the bare JPM as the coupling strength is increased. This is the dark blue band in the 2d plot in Fig. \ref{fig:lscommcalcs}f that shifts from $\sim$1.4 THz to 1 THz as F$_{1}$ is increased. The black horizontal curve is the value of the coupling strength F$_1$ that gives the same lower polariton frequency as observed in experiment (see Fig. 4, discussed below). However, as the coupling strength is increased the upper edge in reflectivity broadens and becomes ill-defined. This is in contrast to the effective medium calculation, where the upper polariton is also seen as a sharp edge splitting from the lower polariton. In short, these calculations reveal that the lower reflectivity edge corresponds to the lower polariton and serves as the observable for the coupling in our MRA-LSCO structure. 

\subsection{Experimental Results and Full Wave Simulations}

\begin{figure*}
    \centering
    \includegraphics[width=17cm]{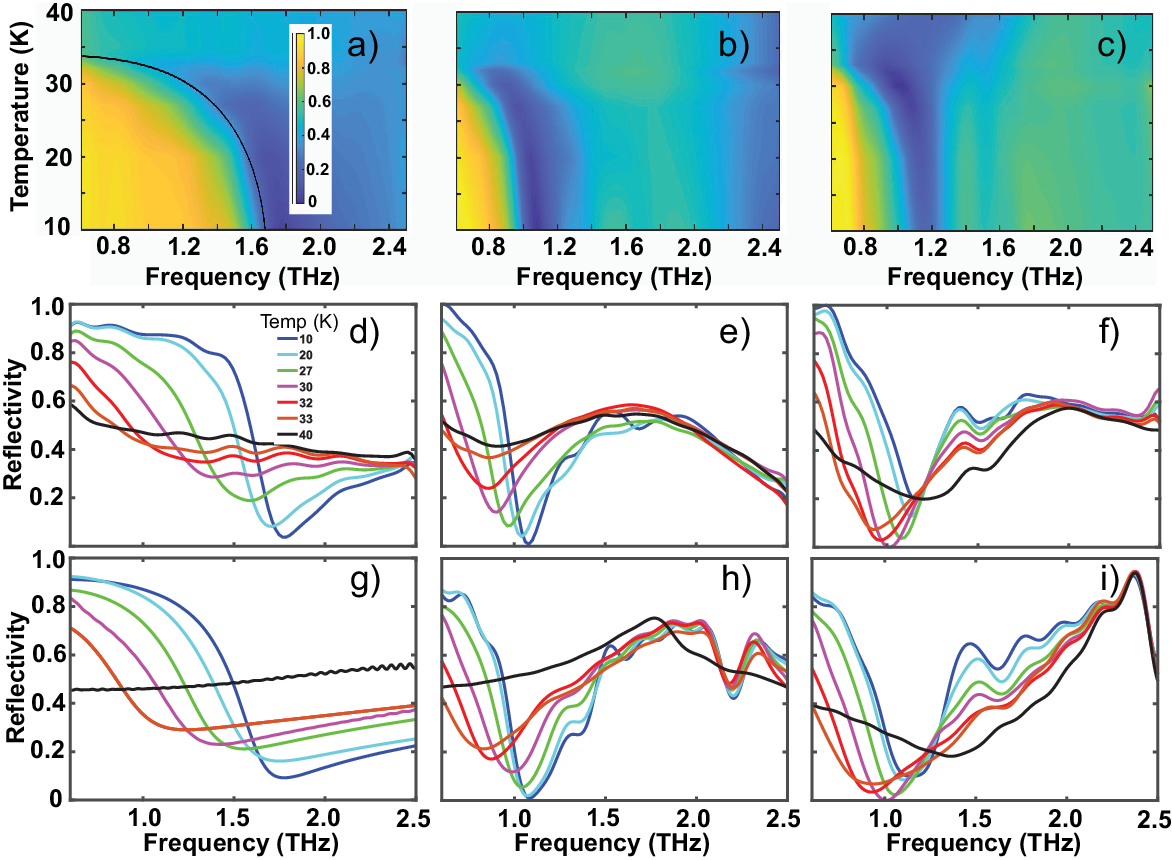}
    \caption[Experimental THz reflectivity and full-wave simulations for MRA-LSCO samples]{Experimental results and full-wave simulations for MRA-LSCO reflectivity. a-c) The 2-d color plots show the trend of the experimental THz reflectivity as a function of temperature for a) bare LSCO crystal, b) 200 nm MM-tape on LSCO and c) 9 $\mu m$ MM-tape on LSCO. d-f) Individual experimental reflectivity curves for d) bare LSCO crystal, e) 200 nm MM-tape on LSCO and f) 9 $\mu m$ MM-tape on LSCO. g-i) Full-wave simulation results for the reflectivity of g) bare LSCO crystal, g) 200 nm MM-tape on LSCO and i) 9 $\mu m$ MM-tape on LSCO.}
    \label{fig:lscommexpt}
\end{figure*}

With the understanding afforded by the simulations and analysis above, we now turn to the experimental data, shown in Fig. \ref{fig:lscommexpt}. For comparison, the reflectivity of the bare LSCO crystal is replotted in Fig. \ref{fig:lscommexpt}a,d. The JPM reflectivity edge redshifts with increasing temperature as the superfluid density decreases. The data is also shown as a 2-d color plot, and the JPM-edge can be seen as the change from gold (high-reflectivity) to blue (low reflectivity). The JPM is indicated by the black curve (Fig. \ref{fig:lscommexpt}a, same fit as in Fig. \ref{fig:lscommfig1}b for clarity). The order parameter behavior can be seen as the gold wedge-like feature of high reflectivity, which becomes smaller as temperature is increased. 

The experimental reflectivity for the MRA-LSCO structure (MRA 200 nm from the LSCO surface) is shown in Fig. \ref{fig:lscommexpt}b,e. A similar edge in the reflectivity ($\sim$1.1 THz at 10 K) is evident and redshifts with increasing temperature. From the discussion and analysis associated with Fig. \ref{fig:lscommcalcs}, it is clear that this is the lower superconducting polariton (SCP) which is  at a highly renormalized frequency compared to the bare JPM. At 10 K, the bare JPM frequency is 1.7 THz while the SCP is at 1.1 THz. This is a shift of $\sim$30$\%$ of the bare JPM frequency. For the MRA-LSCO samples, some of the reflectivity curves had $R$ slightly larger than unity at low frequencies due to imperfect referencing. As such, the 10 K data was normalized to unity, with the higher temperatures then normalized to the 10 K data. The small errors in amplitude do not affect the frequency dependence. The SCP frequency shift of the reflectivity edge with temperature is shown in Fig. \ref{fig:lscommexpt}b. In comparison to the bare crystal, the SCP reflectivity edge occurs at lower frequencies, although showing a similar temperature dependence. 

The reflectivity of Fig. \ref{fig:lscommexpt}e also has a maximum at $\sim$1.6 THz, decreasing gradually at higher frequencies. This is related to the dipole resonance, as seen in the multilayer reflectivity calculated in Fig. \ref{fig:lscommcalcs}d. This maximum in the reflectivity is very broad and is also present at 40 K (black curve of Fig. \ref{fig:lscommexpt}e). This dipole resonance frequency is clearly shifted compared to the bare tape (2.2 THz for the bare tape versus 1.6 THz when the resonators are 200 nm from the LSCO surface). This reflectivity peak does not exhibit a strong temperature dependence (in contrast to lower SCP). This is captured by the analytical calculations of Fig. \ref{fig:lscommcalcs}d. In short, the temperature dependence of the lower SCP is the experimental signature of strong light-matter coupling in this system.  

Figure \ref{fig:lscommexpt}c,f shows the experimental reflectivity for the LSCO-MM sample with the metamaterial array 9 $\mu m$ from the LSCO surface. The lower SCP is again clearly evident and exhibits the same redshift with increasing temperature. The distance between the metamaterial structures and the LSCO surface is varied by a factor of $\sim$40 between the $\sim$9 $\mu m$ and 200 nm tapes. Even when the MRA is further away from the LSCO surface the coupling to the JPM is evident from the shifting of polariton with temperature and the renormalized reflectivity edge ($\sim$1.2 THz at 10 K). As with the 200 nm spacing, the upper polariton is very broad without a strong temperature dependence, in accordance with the analytical multilayer calculations in \ref{fig:lscommcalcs}d. The 9 $\mu m$ reflectivity also clearly shows the peak-dip feature of the asymmetric resonance, at $\sim$1.45 THz for all temperatures below T$_c$. The feature is also at a slightly but noticeably higher frequency at 40 K compared to all temperatures below T$_c$. This is indicative that not only is there a shift of the resonance due to the dielectric constant of LSCO, which is present at 40 K, but also near-field coupling (interlayer capacitance and inductance) which is most pronounced when the LSCO has a c-axis metallic-like response as in the superconducting state \cite{Chen2020}. Finally, we note that the experimental results do not show a change in Tc of the LSCO-MM samples compared to the bare crystal, to within the precision of the measurements of 2K.

The analytical models (Fig. 3) of the MRA-LSCO samples provide some intuition into electromagnetic JPM-MRA coupling but are phenomenological in nature.  As such, full-wave electromagnetic simulations of the LSCO-MM samples were performed to gain further insights. The gold resonator geometry of the MRA is shown in Fig. \ref{fig:lscommfig2}c, with the structural parameters listed in the caption. The 9.4 $\mu m$ thick MRA-polyimide on top of bulk LSCO was simulated with full-wave electromagnetic simulations using the dielectric function extracted from the experimental reflectivity (Fig. \ref{fig:lscommfig1}a). The results are presented in Fig. \ref{fig:lscommexpt}g-i. The simulated bare crystal reflectivity is shown in Fig. \ref{fig:lscommexpt}g. The simulations of the LSCO-MM samples for the 200 nm and 9 $\mu m$ MRAs are shown in panels h and i, respectively. Fig. \ref{fig:lscommexpt}h captures the qualitative features of the 200 nm experimental data, including the lower polariton redshift with increasing temperature, the weak feature of the asymmetric resonance, and the broad hump at higher frequencies. The simulations for the 9 $\mu m$ reflectivity in panel i also show the lower polariton feature and the asymmetric resonance, except for at 40 K (above T$_{c}$). Both the 200 nm and 9 $\mu m$ simulations exhibit an additional sharp dip in the reflectivity that is not present in experiment. This occurs at 2.25 THz for the 200 nm sample and slightly above 2.5 THz for the 9 $\mu m$ sample. Inhomogeneous broadening due to spatial variations in the metamaterial parameters will broaden features in the experimental reflectivity in comparison to simulations, which use a perfect periodic array based on identical unit cells. For example, in comparing the simulated transmission of the bare tape with the experimentally measured transmission in Fig. \ref{fig:lscommfig2}a, a resonance peak at $\sim$2.5 THz in simulations is not present in the experimental transmission. Similarly, the  sharp resonance in the MRA-LSCO simulations is not evident in the experimental measurements of the MRA-LSCO samples. The surface currents in the simulations for both the bare tape and MRA-LSCO samples for this mode show the same behavior, pointing to this being the same mode. As the analysis in the previous section shows, the SCP is mainly related to the coupling between the JPM and the much broader dipole mode of the metamaterial, making it less sensitive to inhomogeneous broadening. Notably, the full-wave simulations match the trends with temperature observed in experiment for the SCP. For example, the depth of the reflectivity minimum of the lower polariton is, for the 200 nm sample, most pronounced at 10 K and increases with increasing temperature. On the other hand the 9 $\mu m$ sample has the lowest reflectivity minimum at 30 K.

\begin{figure}
    \centering
    \includegraphics[width=8.5cm]{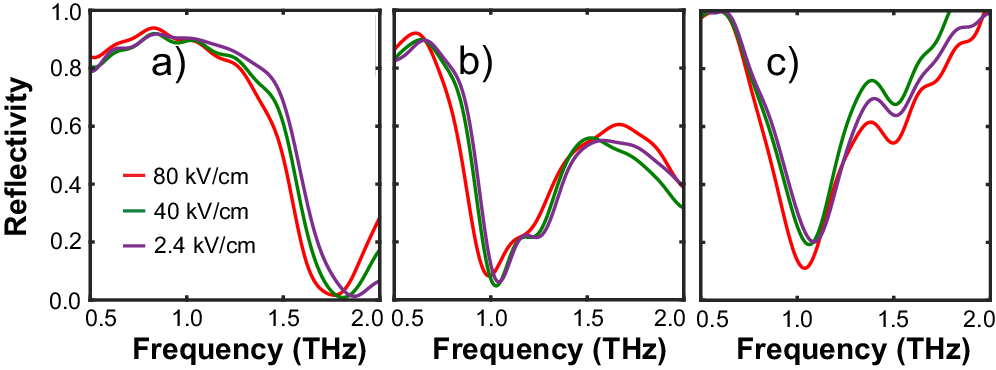}
    \caption[High Field THz Reflectivity]{High-Field THz Reflectivity for a)  bare LSCO crystal, b) 200nm MRA-LSCO structure, and c) 9$\mu$m MRA-LSCO structure.}
    \label{fig:highfield}
\end{figure}

As with any reflectivity plasma edge LSCO can, in principle, operate as an epsilon-near zero (ENZ) material. ENZ materials are of interest as a novel means to achieve linear and nonlinear control of light \cite{Suresh2021,Kinsey2019,Minerbi2022,Litvinov2018,Reshef2019,Jia2021,Jun2023,Vincenti2020,Kumari2023,Habib2020,Alu2007,Wu2021,Yan2023,Basharin2013,Pena2017,Cabezon2020}. In this sense, the temperature dependence of the JPM provides a mean of tuning ENZ response. In addition, LSCO (and other layered cuprates) exhibit nonlinear optical responses at THz frequencies arising from the superconducting response \cite{Kaj2023,Katsumi2023,Rajasekaran2018,Rajasekaran2016,Dienst2011,Dienst2013}. Thus, LSCO is a rather exotic nonlinear ENZ material. Our MRA-LSCO structures provide an additional degree of flexibility to the ENZ response of LSCO arising from polaritonic tuning of the plasma edge. To this end, we performed initial high-field THz reflectivity measurements of the bare LSCO crystal and MRA-LSCO structures as shown in Fig. \ref{fig:highfield}. For the bare LSCO (Fig. 5a), there is a redshift of the JPM frequency with increasing field strength, arising from from field-induced modification of the c-axis tunneling as is evident from a straightforward application of the Josephson equations \cite{Kaj2023,Katsumi2023,Rajasekaran2018}. The high-field THz reflectivity of the MRA-LSCO structures are shown in Fig. \ref{fig:highfield}b(c) for the 200 nm (9$\mu$m) sample. In both cases, the lower polariton redshifts with increasing THz field strength. Beyond the nonlinear tuning of the ENZ response, this further establishes JPM character of the lower polariton edge. THz third harmonic generation (THG) is another well-known nonlinearity for the JPM in cuprates \cite{Kaj2023,Katsumi2023,Rajasekaran2018}. The MRA-LSCO samples do not show enhanced THG compared to the bare crystal (see Ref. \cite{Kaj2023} for details). The main reason for this is that the metamaterial dipole mode is close to the THG frequency, so the low transmission of the tape would effectively block the transmission of generated third harmonic radiation. Further studies optimizing the field enhancement and the frequencies of the metamaterial modes and polaritons with respect to the JPM and the THz drive give the possiblity of enhanced THG and nonlinear manipulation of the JPM-MRA coupling and corresponding ENZ response.

\section{Conclusions}
The modeling presented above demonstrates that the large frequency shifts of the effective longitudinal response of our MRA-LSCO sample arise from  classical electromagnetic coupling.  The observed response does not correspond to a fundamental change to the quantum phase, such as a change to in the superfluid density or critical temperature. Nonetheless, these results are important towards providing an understanding of the electromagnetic effects that arise in searching for cavity mediated superconductivity and highlight the need for a careful electromagnetic analysis prior to any interpretation using quantum based models. Polaritonic shifts are insufficient to identify changes the quantum properties of a material.
In other words, while a large Rabi-splitting is indicative of strong or ultra-strong light matter coupling, a change to the properties of the quantum phase or other non-classical-electromagentic effects such as radiative transitions or vacuum Bloch-Siegert Shift are needed to show effects of vacuum fluctuations or quantum coupling is achieved \cite{Cong2016,Bamba2022,Li2018,FornDiaz2019}. 

In this work electromagnetic coupling between metamaterial resonator arrays and the Josephson Plasma Mode (JPM) in LSCO was investigated using THz reflection spectroscopy. A clear coupling manifests as a renormalized plasma edge corresponing to the lower superconducting polariton (SCP). Models were developed to explain how the lower SCP seen in reflectivity is fully indicative of the strong coupling of the system, and that the edges in reflectivity show a shift to the longitudinal response of the system. Our metamaterial tapes allow the MRA-LSCO distance to be varied and reveal stronger coupling when the MRA is closer to the LSCO surface. This work experimentally demonstrates coupling between a cavity-like MM resonance and the collective mode of a quantum material, and identifies the observables for coupling to the c-axis response of the superconductor. Our results also demonstrate the ability explore light-matter-cavity coupling using a replaceable metamaterial structure as the cavity, allowing multiple cavity-resonator-structures to be used with the same single-crystal sample. Further work and design optimization could result in stronger coupling thereby modifying the superconducting state, or realizing nontrivial, nonlinear light-matter coupling effects, such as Dicke Superradiance, vacuum Bloch-Siegert Shift, or cavity-mediated squeezing \cite{Cong2016,Li2018,Hayashida2023,Bamba2022}. However, considerable care will be required to extract signatures that correspond to modification of the quantum phase beyond conventional electrodynamic coupling.

Acknowledgements: Research at UCSD supported by the U.S. National Science Foundation DMR-1810310 and NASA 80NSSC19K1210. Research at BU is supported by U.S. National Science Foundation ECCS-1810252. Research at Columbia is supported by DMR-2210186 and  DMR-2011738. DNB is is the Vannevar Bush Faculty Fellow ONR-VB: N00014-19-1-2630.

\bibliographystyle{ieeetr}
\bibliography{library}

\end{document}